\documentclass[12pt]{JHEP3}

\usepackage{epsfig}
\epsfclipon
\usepackage{multicol}
\usepackage{epsfig,bm}
\usepackage{amssymb,amsmath}
\newcommand{\roughly}[1]{\mathrel{\raise.3ex\hbox{$#1$\kern-0.85em
\lower1ex\hbox{$\sim$}}}}
\newcommand{\lsim}{\roughly<}
\newcommand{\gsim}{\roughly>}

\def\cA{{\cal A}}
\def\cB{{\cal B}}

\def\cH{{\cal H}}

\def\cU{{\cal U}}
\def\cV{{\cal V}}

\def\Dsl{\hbox{/\kern-.6700em\it D}} 
\def\dsl{\hbox{/\kern-.5300em$\partial$}}

\def\eqa{\begin{eqnarray}}
\def\eeqa{\end{eqnarray}}
\def\eq{\begin{equation}}
\def\eeq{\end{equation}}
\def\be{\begin{equation}}
\def\ee{\end{equation}}
\def\bea{\begin{eqnarray}}
\def\eea{\end{eqnarray}}
\def\nn{\nonumber}

\def\pref#1{(\ref{#1})}

\def\Tr{{\rm Tr \,}}
\def\exd{{\rm d}}

\title{On the Decoherence of Primordial Fluctuations During Inflation}
\author{C.P. Burgess,$^{1,2}$ R.
Holman${}^{3}$ and D. Hoover$^{4}$ \\
$^1$ Department of Physics and Astronomy, McMaster University,\\
\qquad Hamilton, ON, Canada, L8S 4M1\\
$^2$ Perimeter Institute, Waterloo, ON, Canada N2L 2Y5\\
$^3$ Physics Department, Carnegie Mellon University, Pittsburgh PA
15213 USA\\
$^4$ Dept. of Physics, McGill University, 3600 University
Street,\\
\qquad Montr\'eal QC, Canada H3A 2T8}

\abstract{We study the process whereby quantum cosmological
perturbations become classical within inflationary cosmology. By
setting up a master-equation formulation we show how quantum
coherence for super-Hubble modes can be destroyed by its coupling
to the environment provided by sub-Hubble modes. We identify what
features the sub-Hubble environment must have in order to decohere
the longer wavelengths, and identify how the onset of decoherence
(and how long it takes) depends on the properties of the
sub-Hubble physics which forms the environment. Our results show
that the decoherence process is largely insensitive to the details
of the coupling between the sub- and super-Hubble scales. They also show how
locality implies , quite generally, that the decohered density matrix
at late times is diagonal in the field representation (as is
implicitly assumed by extant calculations of inflationary density
perturbations). Our calculations also imply that decoherence can
arise even for couplings which are as weak as gravitational in
strength. }

\begin{document}

\makeatletter \@addtoreset{equation}{section} \makeatother
\renewcommand{\theequation}{\thesection.\arabic{equation}}

\setcounter{page}{1} \pagestyle{plain}
\renewcommand{\thefootnote}{\arabic{footnote}}
\setcounter{footnote}{0}

\section{Introduction and Summary}

Recent precision observations \cite{WMAP,boomerang,dasi} of
temperature fluctuations within the Cosmic Microwave Background
(CMB) agree well \cite{WMAPInflation} with the predictions made
for them by inflationary models, according to which the Universe
once underwent an accelerated expansion during its remote past
\cite{Inflation}. According to the inflationary picture, the
observed CMB temperature variations are seeded by tiny primordial
density perturbations in the much earlier universe, perturbations
which began as quantum fluctuations during the earlier
inflationary epoch.

More precisely, within the inflationary picture the primordial
density fluctuations derive from the quantum fluctuations,
$\langle \phi(x) \phi(y) \rangle$, of the quantum inflaton field,
$\phi$, whose dynamics drives the signature accelerated expansion.
The success of the inflationary picture requires these {\it
quantum} fluctuations to be converted into classical {\it spatial}
variations in the energy density whose amplitude varies
stochastically as one moves from one Hubble patch to another
within the Universe during the much later recombination epoch.

At present, the classicalization of primordial perturbations is
understood in terms of the properties of the quantum state into
which the fluctuation fields evolve (see, for example
\cite{Guth:1985ya,Polarski:1995jg}). It is known that both scalar
and tensor density fluctuations evolve during inflation into a
particular kind of quantum state -- one very similar to the
squeezed state of quantum optics
\cite{Grishchuk:1990bj,Albrecht:1992kf}. In particular, quantum
expectation values of products of fields in a highly squeezed
state turn out to be identical to stochastic averages calculated
from a stochastic distribution of classical field configurations,
up to corrections which vanish in the limit of infinite squeezing
\cite{Polarski:1995jg,Kiefer:1998qe,Lesgourgues:1996jc}. Quantum
autocorrelation functions like $\langle \phi(x) \phi(y) \rangle$
become indistinguishable from autocorrelations for $c$-number
fields $\varphi(x)$ computed within a classical stochastic
process, when computed in the limit of large squeezing.

Our purpose in this note is to describe what these analyses do
{\it not} address: the decoherence process through which the
initial quantum state evolves into the corresponding stochastic
ensemble of field configurations. That is, we wish to understand
how the density matrix, $\rho[\varphi(x),\varphi'(y)] = \langle
\varphi(x)| \rho | \varphi'(y) \rangle$, for each of the various
quantum fields, $\phi(x)$, evolves from an initially pure state,
\be \label{PureState}
    \rho[\varphi(x),\varphi'(y)] = \Psi[\varphi(x)]
    \Psi^*[\varphi'(y)] \,,
\ee
describing the initial conditions (where $\Psi[\varphi(x)] =
\langle \varphi(x)|\psi \rangle$), to a mixed state,
\be \label{DecoheredState}
    \rho[\varphi(x), \varphi'(y)] = P[\varphi(x)] \,
    \delta[\varphi(x) - \varphi'(y)] \,,
\ee
describing the later classical stochastic probability distribution
for the field amplitudes. Such a transition is implicit in the
standard predictions of inflationary implications for the CMB, and
our goal is to understand to what extent these predictions might
depend on the details of how this decoherence is achieved.

Decoherence is a much-studied process outside of cosmology,
although not all of the conceptual issues have been resolved. In
particular, the transition from pure to mixed state we seek
typically arises whenever the degrees of freedom of interest
interact with an `environment' consisting of other degrees of
freedom whose properties are not measured. The measured degrees of
freedom can make a transition from a pure to a mixed state once a
partial trace is taken over this environmental sector.

We are most interested in those modes which are responsible
for the correlations which are observed in the CMB and in
structure formation -- modes which have only come within the
Hubble scale relatively recently. We take the relevant decohering
environment to consist of those modes having much shorter
wavelength, which entered the Hubble scale much earlier and whose
correlations are not directly visible in the sky. By borrowing
techniques from atomic and condensed-matter physics we set up a
master equation for the density matrix of the observed modes which
allows us to robustly identify how these modes decohere provided
only that ($i$) the environment is large enough to not be overly
perturbed by the modes whose decoherence we follow; ($ii$) the
interactions with the environment are weak; and ($ii$) the
characteristic correlation time of fluctuations in the environment
is much smaller than the timescales of interest to the decoherence
process.

We use this formalism to identify how the decoherence depends on
the properties of the short-wavelength modes which are assumed to
make up the environment. Other authors have also examined the
decoherence problem, often using the formalism of influence
functionals \cite{Feynman} (see, for example
\cite{Kiefer:1998qe,Sakagami:1987mp,Brandenberger:1990bx,
Calzetta:1995ys,Lombardo:2005iz,Grishchuk:1989ss}).
Unfortunately, the complexity of this formulation often forces
calculations to be performed only for free fields, which couple to
one another only by mixing in their mass or kinetic terms, leaving
the uncertainty as to whether the results are restricted to the
special choices which are made for calculational purposes. The
main virtue of our treatment in terms of the master equation is
that it allows us to identify the nature of the decoherence
process in a controllable approximation which can encompass very
general, realistic environment-system interactions.

Our analysis leads to the following results:
\begin{itemize}
\item We find that the quantity in the environment which controls
the decoherence process is the autocorrelation function of the
interaction Hamiltonian, $V(t)$, which couples the observed modes
to the environment: $\Tr_{\rm env} [\rho \,\delta V(t) \, \delta
V(t')]$, where the trace is over the environmental sector, $\rho$
denotes the system's density matrix and $\delta V(t) = V(t) -
\Tr_{\rm env}[\rho \, V(t)]$ represents the fluctuations of $V(t)$
(in the interaction representation). More complicated dependence
on the properties of the environment only arise at higher order in
$V(t)$.
\item We find that the generic situation is that the observed modes
decohere into a density matrix of the form of
eq.~\pref{DecoheredState}, which is diagonal when expressed in
terms of field eigenstates. This basis typically plays a special
role because locality usually dictates that the important
interactions between the modes of interest and their environment
are diagonal when written in this basis. This point has previously
been emphasized in ref.~\cite{Kiefer:1998jk}
\item We show that the final stochastic probability distribution
in eq.~\pref{DecoheredState} is given in terms of the initial pure
state by $P[\varphi(x)] = |\langle \varphi(x)|\psi\rangle|^2$, and
thereby justify more precisely the standard arguments which
implicitly use the classical stochastic field distribution which
reproduces the computed initial quantum fluctuations of the
inflaton.
\item We find that short wavelength modes of the environment have
the potential to decohere longer-wavelength modes, but need not do
so. In particular they should not do so if the environment is
prepared with its modes in their adiabatic vacua. It is this
observation which explains why we are able to measure quantum
states at all in the lab, given that there are always an enormous
number of short-wavelength modes to which any given experiment
does not have access.
\item We find that for simple choices for the short-distance
environment there is ample time for decoherence to occur during or
after inflation, even if the modes of interest are coupled to the
environment with gravitational strength. Since all modes couple
gravitationally, this shows that decoherence is ubiquitous for
modes produced during inflation which are now observable on the
largest scales. In the special case that decoherence arises due to
Planck-suppressed couplings between super-Hubble inflaton modes
and thermal sub-Hubble modes after reheating, decoherence is only now
beginning to occur for those modes which cross the horizon during
the recombination epoch, raising the intriguing possibility that
this might have observable implications.
\end{itemize}

We next present these results in more detail. We do so by first
briefly reviewing our approach to decoherence with an emphasis on
the limits of validity of our calculations. We then examine two
candidate environments, both of which are known to exist in most
inflationary scenarios: ($i$) the short-wavelength modes of the
metric and inflaton themselves, during inflation and afterwards
(for which we find our calculations break down, although we give
arguments as to why decoherence from this source may be unlikely);
and ($ii$) those short-wavelength modes which are the first to
thermalize after inflation ends (for which we argue our
calculations apply, and show that decoherence could reasonably
occur in the time available).

Our purpose is not to argue that either of these sources must
provide the dominant source of decoherence, but rather to argue
that our calculational tools suffice to identify that environments
exist which have sufficient time to decohere initial quantum
fluctuations. Our hope is that these tools can be used to compare
potential sources of decoherence, with a view to identifying those
which are the most important and how the detailed features of
decoherence depend on the nature of the environment assumed. In
particular, we hope to find observational consequences of the
decoherence process which might be possible to use to test further
the inflationary paradigm for generating primordial density
fluctuations. We group some of the details of the calculations
into appendices.

\section{Decoherence and Environments}

The word `decoherence' can mean a variety of things (see, for
example, \cite{Zeh:1999fs,Zurek} for extensive discussions), but
for the present purposes we intend it to refer to the transition
from a state which is initially pure --- {\it i.e.} its density
matrix can be written $\rho = |\psi \rangle \langle \psi |$, for
some state-vector $|\psi \rangle$ --- to a mixed state (for which
no such state-vector exists). Our interest is in how long this
transition takes, and in the form of the final mixed state to
which the system is led. Once the final mixed state is obtained,
its interpretation as a classical statistical ensemble is most
simply found using the basis in which it is diagonal, since in
this basis we have $\rho_{\rm red} = \Tr_{\rm env} \rho = \sum_n
p_n | n \rangle \langle n |$, with $0 \le p_n \le 1$ denoting the
classical probability of finding the system in state $|n\rangle$
\cite{Zurek2}.

Convenient diagnostics for determining the purity of the state
represented by a density matrix $\rho$ are given by the
basis-independent expressions: $\rho^2 = \rho$ and $\Tr[\rho^2] =
1$, either of which are satisfied if and only if the state
described by $\rho$ is pure. The equivalence of these criteria for
purity rely on the fact that density matrices are defined to be
hermitian, positive semi-definite and satisfy $\Tr[\rho] = 1$.

\subsection{The Master Equation}

We now set up a master equation which allows us to describe the
pure-to-mixed transition quantitatively, and in a way which can be
applied to a broad variety of choices for both the observable
sector and the environment. Our presentation here follows closely
that of ref.~\cite{BM} (see also \cite{Feynman,others} for
alternative approaches). To do so we write the system's Hilbert
space to be the direct product of the observable sector, $A$, and
the environment, $B$: ${\cal H} = {\cal H}_A \otimes {\cal H}_B$,
and write the total Hamiltonian governing the system as the sum $H
= H_A + H_B + V$, where $H_A$ and $H_B$ describe the dynamics of
the subspaces $A$ and $B$ in the absence of their mutual
interactions, and $V$ is the interaction term which couples them.

For the applications of present interest we may also imagine the
system starts off without any correlations between these two
sectors, and so is described by an initial density matrix: $\rho(t
= t_{0}) = \varrho_A \otimes \varrho_B$. The presence of the
interaction $V$ generates correlations between the two sectors as
the system evolves; these correlations make a general description
of their further evolution difficult. A great simplification is
possible, however, if three conditions are satisfied
\cite{API,NOptics,formalism}:
\begin{enumerate}
\item
The interactions between sectors $A$ and $B$ are weak;
\item
System $B$ is large enough not to be appreciably perturbed by its
interactions with system $A$;
\item
The correlation time, $\tau$, over which the autocorrelations
$\langle \delta V(t) \, \delta V(t - \tau) \rangle_B$ are
appreciably nonzero for the fluctuations of $V$ in sector $B$, are
sufficiently short, $V\tau \ll 1$.
\end{enumerate}
In this case the dynamics of system $A$ over times $t \gg \tau$
does not retain any memory of the correlations with $B$ since
these only survive for much shorter times. This allows us to treat the
evolution of $A$ in the presence of $B$ as a Markov
process. What emerges is a picture wherein the interactions with
$B$ simply provide a stochastic component to the evolution of
sector $A$. From this picture, we can derive a master equation which
describes how the environment $B$ affects the evolution of the
density matrix describing sector $A$, and in particular how
quickly sector $B$ leads to decoherence in system $A$.

It is convenient to work within the interaction representation,
for which the time evolution of operators is governed by $H_A +
H_B$ while the evolution of states is governed by $V$. We
concentrate on obtaining an evolution equation for the reduced
density matrix, $\rho_A(t) = \Tr_B[\rho(t)]$, since this includes
all of the information concerning measurements involving
observables only in sector $A$. As Appendix \ref{App: ABFormalism}
shows, writing the interaction Hamiltonian in terms of product
basis of operators --- $V = \sum_j A_j \otimes B_j$ --- and using
the three assumptions given above allows the derivation of the
following master equation for $\rho_A(t)$:
\be \label{masterequationbody}
    \frac{\partial \rho_A}{\partial t} = i \Bigl[\rho_A
    , A_j \Bigr] \langle B_j \rangle_B
    -\, \frac12 \, {\cal W}_{jk}  \Bigl\{ \Bigl(
    \rho_A A_j A_k + A_j A_k \rho_A - 2 A_k \rho_A A_j
    \Bigr) \Bigr\} \,.
\ee
where there is an implied sum over any repeated indices, and
$\langle \cdots \rangle_B = \Tr_B[(\cdots)\varrho_B]$ denotes the
average over the initial (and unchanging) configuration of sector
$B$. The assumption of a short correlation time, $\tau$, has been
used here to write
\be \label{shorttimecorrs}
    \langle \delta B_j(t) \delta B_k(t') \rangle_B = {\cal
    W}_{jk}(t) \delta(t-t') \,,
\ee
which defines the $O(\tau)$ coefficients ${\cal W}_{jk}^* = {\cal
W}_{kj}$.

The crucial observation is that the three assumptions listed above
ensure that corrections to this equation are of order $(V\tau)^3$,
where $\tau$ denotes the correlation time over which quantities
like $\langle \delta V(t) \, \delta V(t + \tau) \rangle_B$ are
appreciably nonzero. The use of perturbation theory to evaluate
$\partial \rho_A/\partial t$ therefore requires only that $\tau$
be sufficiently short: $V\tau \ll 1$. However once
eq.~\pref{masterequationbody} is justified in this way, its
solutions can be trusted even if they are integrated over times
$t$ for which $Vt \gg 1$.

\subsection{Local Interactions and
Spatially-Small Fluctuations}

We now specialize to the case of interest for field theory, where
interactions are local. Suppose, then, that sector $A$ and $B$
interact through interactions of the local form:
\be \label{Vdef}
    V(t) = \int d^3x \; \cA_i(x,t) \, \cB_i(x,t) \,,
\ee
where there is an implied sum on $i$, ${\cal A}_i$ denotes a local
functional of the fields describing the $A$ degrees of freedom,
and ${\cal B}_i$ plays a similar role for sector $B$.

In the special case that the degrees of freedom in sector $B$ have
very short correlation lengths, $\ell$, (as well as times,
$\tau$), their influence on $\partial \rho_A/\partial t$ can be
represented in terms of local `effective' interactions. That is,
suppose
\be \label{shortdistcorrs}
    \langle \delta {\cal B}_i(x,t) \, \delta {\cal B}_j(x',t')
    \rangle_B = \cU_{ij}(x,t) \, \delta^3(x-x') \, \delta(t-t') \,,
\ee
where $\cU_{ij}^* = {\cal U}_{ji}$ is a calculable local function
of position which is of order $\ell^3 \tau$. As is shown in
Appendix \ref{App: ABFormalism}, under this assumption the master
equation, eq.~\pref{masterequationbody} becomes:
\be \label{dmeqbody}
    \frac{\partial \rho_A}{\partial t} =
    i  \int
    d^3x \; \Bigl[ \rho_A , \cA_j
    \Bigr] \, \langle \cB_j
    \rangle_B
    -  \frac12 \int d^3x \; \cU_{jk} \Bigl[
    \cA_j \cA_k \rho_A + \rho_A \cA_j \cA_k
    - 2 \cA_k \rho_A \cA_j \Bigr]  \,.
\ee
Notice that this equation trivially implies $\partial
\rho_A/\partial t = 0$ for any $\rho_A$ which commutes with all of
the $\cA_j$. Eqs.~\pref{masterequationbody} and \pref{dmeqbody}
are the main results in this section, whose implications we
explore in detail throughout the remainder of the paper.

We remark that eqs.~\pref{shorttimecorrs} and
\pref{shortdistcorrs} are a fairly strong conditions, inasmuch as
they exclude significant correlations along the entire light cone.
In particular, they exclude the application of the master
equations, \pref{masterequationbody} and \pref{dmeqbody}, to
situations where $\langle \delta {\cal B}_j(x,t) \delta {\cal
B}_k(x',t') \rangle$ depends only on the proper separation,
$s(x,t;x',t')$, between $(x,t)$ and $(x',t')$, such as when $B$ is
prepared in a Lorentz-invariant vacuum (like the Bunch-Davies
vacuum of de Sitter space).

\subsection{Solutions}

If all of the operators $\cA_i(x,t)$ commute at equal times then
the implications of eq.~\pref{dmeqbody} are most easily displayed
by taking its matrix elements in a basis of states $|\alpha
\rangle_t$ for which the operators $\cA_i(x,t)$ are diagonal:
$\cA_i(x,t) |\alpha \rangle_t = \alpha_i(x) | \alpha \rangle_t$
(for fixed $t$). With this choice we denote
\be
    \rho_t[\alpha,\tilde\alpha] =  {}_t\langle \alpha| \rho_A|
    \tilde\alpha
    \rangle_t,
\ee

and so using the interaction-picture evolution $\partial_t |\alpha
\rangle_t = -iH_A |\alpha\rangle_t$, we find
\bea \label{drhodteqn}
    \frac{\partial\rho_t}{\partial t}[\alpha,\tilde\alpha]
    &=& (\partial_t \langle \alpha |) \rho_A |\tilde\alpha \rangle
    + \langle \alpha | \partial_t \rho_A |\tilde\alpha \rangle
    + \langle \alpha | \rho_A (\partial_t |\tilde\alpha
    \rangle)\nonumber\\
    &=& \langle \alpha | \partial_t \rho_A |\tilde\alpha \rangle
    + i \langle \alpha | [\rho_A, H_A] |\tilde\alpha \rangle \,.
\eea
Notice that the last term in this equation combines with the term
involving $\langle V \rangle_B$ in eq.~\pref{dmeqbody} to give the
combination $i \langle \alpha | [ \rho_A, H_A + \langle V
\rangle_B] | \tilde\alpha \rangle$ in the evolution of the matrix
element $\rho_t[\alpha,\tilde\alpha]$. Because these terms describe
a Hamiltonian evolution they cannot generate a mixed state from
one which is initially pure. For our present purposes of
understanding decoherence we need follow only those terms of
eq.~\pref{drhodteqn} which are $O(V^2)$ and which do not simply
describe pure-state evolution.

The general solution to this equation is easily given in those
circumstances for which the term involving $H_A$ is negligible,
since in this case evaluating the matrix elements gives
\bea \label{dmeqscalarcosmologyme}
    &&\frac{\partial \rho_t}{\partial t} =
    -i \rho_t
    \int d^3z  \Bigl[ \alpha_i(z) - \tilde\alpha_i(z)
    \Bigr] \langle \cB_i(z,t) \rangle_B \nonumber\\
    && \qquad\qquad\qquad - \rho_t \int d^3z \Bigl[
    \alpha_i(z) - \tilde\alpha_i(z)
    \Bigr]\Bigl[ \alpha_j(z) - \tilde\alpha_j(z) \Bigr] \, \cU_{ij}(z,t) \,.
    \nonumber
\eea
These equations are readily integrated to yield
\be
    \rho_t[\alpha,\tilde\alpha] = \rho_{t_0}[\alpha,\tilde\alpha]
    \, e^{-\Gamma - i \Sigma} \,,
\ee
where
\bea \label{GaussianForm}
    \Sigma &=& \int_{t_0}^t dt' d^3z \Bigl[ \alpha_i(z)  - \tilde\alpha_i(z)
    \Bigr] \langle \cB_i(z,t') \rangle_B \\
    \Gamma &=& \int_{t_0}^t dt'  d^3z \Bigl[
    \alpha_i(z) - \tilde\alpha_i(z) \Bigr]\Bigl[\alpha_j(z)
    - \tilde\alpha_j(z) \Bigr]
    \cU_{ij}(z,t') \,. \nonumber
\eea
This solution describes the evolution of the reduced density
matrix over timescales for which the unperturbed evolution due to
$H_A$ is negligible.

As expected, the mean-field term (involving $\langle \cB_i
\rangle_B$) only changes the phase of the initial pure state and
so it may be neglected to the extent that only the transition from
pure state to mixed is of interest. The same would be true for the
$H_A$ terms in eq.~\pref{drhodteqn} if $[H_A,A_i] = 0$. Otherwise
$H_A + \langle V \rangle_B$ acts to evolve, but not to decohere,
the initial pure state. By contrast, those terms involving the
fluctuations $\cU_{ij}$ cause the reduced density matrix to take
the form of a classical Gaussian distribution in the
$\{\alpha_i(x) \}$, whose time-dependent width is controlled by
the local autocorrelation function $\cU_{ij}$. Provided this width
shrinks in time, at late times the system evolves towards a
density matrix which is diagonal in the $|\alpha \rangle$ basis,
with diagonal probabilities that are time-independent and set by
the initial wave-function:
\be
    P_t[\alpha] \equiv \rho_t[\alpha,\alpha]
    = \rho_{t_0}[\alpha,\alpha] = \Bigl| \Psi_{t_0}[\alpha]
    \Bigr|^2 \,.
\ee
Here $\Psi_{t_0}[\alpha]$ is the wave-functional for the initial
pure state.

These solutions describe the decoherence of the initial state into
the classical stochastic ensemble for the variables, $\{ \alpha_i
\}$, which diagonalize the interactions with the decohering
environment. Since general considerations of locality typically
require the interactions, $\cA_i$ to be ultra-local polynomials of
the fields, this shows that the classical distribution to which
decoherence leads is generically an ensemble in field space,
$\varphi(x)$ \cite{Kiefer:1998jk,Zurek2}, just as is implicitly
assumed in standard calculations of inflationary perturbations.

\section{Decoherence in Cosmology}

We next apply the above formalism to inflationary cosmology, in
order to identify potential sources of decoherence for those
fluctuations whose influence is observed in CMB measurements.

\subsection{Generalization to FRW Geometries}

We start with a short discussion of how eq.~\pref{dmeqbody} must
be modified when applied to an expanding Friedmann, Robertson
Walker (FRW) universe. We may do so either using co-moving time or
conformal time, leading to physically equivalent results.

The starting point is the generalization of eq.~\pref{Vdef} to
curved space. If we take the interactions $\cA_i$ and $\cB_i$ to
be Lorentz scalars, the insertion of a factor of the 3-metric's
volume element gives
\be \label{Vdef2}
    V(t) = a^3(t) \int d^3x \; \cA_i(x,t) \, \cB_i(x,t) \,.
\ee
Keeping in mind that the delta function transforms as a tensor
density, eq.~\pref{shortdistcorrs} becomes
\be \label{shortdistcorrs2}
    \langle \delta {\cal B}_i(x,t) \, \delta {\cal B}_j(x',t')
    \rangle_B = \left[ \frac{\cU_{ij}(x,t)}{a^{3}(t)} \right]
    \, \delta^3(x-x') \, \delta(t-t') \,,
\ee
where $\cU(x,t)$ is also a Lorentz scalar. With these choices
eq.~\pref{dmeqbody} generalizes to:
\bea \label{dmeqbody2}
    &&\frac{\partial \rho_A}{\partial t} =
    i  a^3 \int
    d^3x \; \Bigl[\rho_A ,  \cA_i
    \Bigr] \, \langle \cB_i
    \rangle_B  \nonumber\\
    &&\qquad  -  \frac{a^3}{2} \int d^3x \; \cU_{jk} \Bigl[
    \cA_j \cA_k \rho_A + \rho_A \cA_j \cA_k
    - 2 \cA_k \rho_A \cA_j \Bigr] \,. \nonumber\\
\eea
The equivalent result in conformal time is obtained using the
replacement $\exd t = a \, \exd \eta$.

\subsection{Short-Wavelength Modes as Environment}

For applications to inflationary predictions for the CMB we take
sector $A$ to be comprised of those modes of the inflaton-metric
system whose wavelength allows them to be visible within current
measurements. In particular, we take $A$ to consist of the
Mukhanov field \cite{Mukhanov2,Lukash},
\be \label{mukhvarbody}
    v =  a\, \left[ \delta\phi +
    \frac{\varphi^{\prime}}{\cH}\psi \right] \,,
\ee
which describes the physically relevant combination of the
inflaton fluctuation, $\delta\phi$, and the scalar metric
fluctuation, $\psi$ (see Appendix \ref{App:MIPerturbations} for
more specific definitions). Here $a(\eta)$ denotes the background
scale factor as a function of conformal time, $\eta$, for which
${\cal H} = a'/a$, and $\varphi$ is the background inflaton
configuration. Primes denote differentiation with respect to
$\eta$. What makes this variable convenient is that its linearized
evolution is governed by a flat-space, free-field action in the
presence of a time-dependent mass, $m^2 = -z''/z$:
\be \label{S_2body}
    S_2\,=\,\frac{1}{2} \int{d^{4}x \, \left[ v^{\prime
    2}- \gamma^{ij} \partial_i v \, \partial_j v+
    \left( \frac{z^{\prime\prime}}{z}
    \right) v^{2} \right]},
\ee
where $z = a\varphi^{\prime}/\cH$ and $\gamma_{ij}$ is the spatial
FRW metric (which we choose in what follows to be flat).

We must then make a choice for the decohering environment, sector
$B$. A minimal choice for $B$ consists of the short-wavelength
modes of the inflaton and metric fields themselves, since these
modes and their mutual interactions are in any case required to
exist by the assumed fluctuation-generating mechanism. In this
section we explore the extent to which these interactions can
suffice to decohere longer-distance inflaton fluctuations, given
their very weak strength.

In particular we ask how the time-scale for decoherence depends on
the physical parameters of this environment, in order to determine
if sufficient time is available for decoherence to have been
completed for those modes relevant to the CMB between horizon exit
and re-entry. Our goal in so doing is not to argue that these are
the most important sources of cosmological decoherence, but rather
to show that the inflaton couples with sufficient strength in the
early universe to plausibly have had enough time to have decohered
before horizon re-entry.

\subsubsection*{Causality Issues}

We have seen that the master equation derived above only applies
if the environment's correlation time, $\tau$, is much smaller
than the times over which the system of interest evolves. Since
the Hubble scale, $H$, sets the natural timescale for evolution in
an expanding universe, this formalism is most naturally applied to
situations where it is sub-Hubble modes ($k/a \gg H$) which
decohere super-Hubble modes ($k/a \lsim H$). Can this type of
decoherence be consistent with causality?

At first sight one might think not, since there are general
arguments \cite{traschen} which restrict how much short-wavelength
quantum fluctuations can modify background fields like $\langle
\phi \rangle = \varphi$. However our interest in this paper is in
how the system's density matrix evolves between horizon exit and
re-entry, and in particular how the transition occurs from a pure
state to one that is mixed. But this is to do with the
entanglement of short- and long-wavelength modes and the
destruction of the phase coherence amongst the long-wavelength
modes which arises once the short-distance modes are traced out.
Since mean quantities like $\varphi$ are not changed by
this decoherence, it is not constrained by considerations such as
those of \cite{traschen}. We believe our situation is closer
to Einstein-Podolsky-Rosen experiments, for which entanglements
between widely-separated electrons can give rise to seemingly
nonlocal evolution of the system's density matrix.

\subsection{Decoherence After Reheating}

\begin{figure}
\begin{center}
\includegraphics[width=0.4\textwidth]{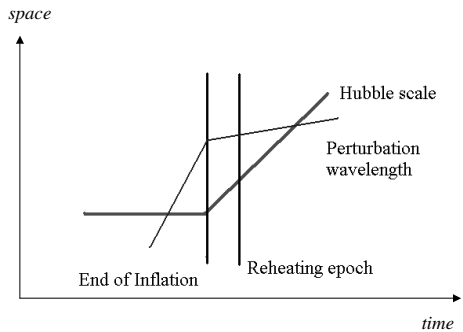}
\caption{A cartoon of the evolution of the wavelength of a
perturbation and the Hubble length vs time during inflation and
afterwards. The vertical lines indicate the end of inflation, and
the beginning of the subsequent reheating epoch.}
\label{Fig:Cartoon}
\end{center}
\end{figure}

We now must decide when to look for decoherence between the epoch
when the observed modes leave the horizon during inflation, and
their re-entry in the not-so-distant past (see Figure
\ref{Fig:Cartoon}). We seek an instance when the short-wavelength
modes are prepared in states for which the short-distance
assumptions, eqs.~\pref{shorttimecorrs} and \pref{shortdistcorrs},
are satisfied.

The simplest such instance occurs after the reheating epoch.
Reheating occurs once inflation ends and modes begin to re-enter
the Hubble scale, since they can then acquire more complicated
dynamics. Detailed studies indicate that under some circumstances
this dynamics can lead to the reheating \cite{Reheating} and/or
preheating \cite{Preheating} of the universe, giving rise to the
thermal state which late-time Big Bang cosmology assumes.

Since thermal fluctuations can often satisfy the short-correlation
assumptions, eqs.~\pref{shorttimecorrs} and \pref{shortdistcorrs},
we now examine whether a thermal bath of sub-Hubble modes can
decohere the longer-wavelength, super-Hubble modes through their
weak mutual self-interactions. Within this picture the first modes
to thermalize would do so as in standard treatments, about which
we have nothing new to say. Our interest is instead in how this
nascent thermal (or other) mixed state acts to decohere those
modes which only enter the Hubble scale much later, whose imprint
we now see in the CMB. Since this type of decoherence cannot begin
until after inflation ends, a key question is whether there is
sufficient time over which to decohere the super-Hubble inflaton
modes before they re-enter.

\subsubsection*{Decoherence Rates}

In order to estimate whether there is sufficient time to decohere
super-Hubble modes we require a simple model for the
lowest-dimension inflaton/heat-bath interaction, which we take to
have the form of eq.~\pref{Vdef}, with $\cA = g v$ and where $\cB$
is a local operator having engineering dimension (mass)${}^d$,
with $d > 0$. On dimensional grounds the coupling $g$ then has
dimension (mass)$^{3-d}$. With these assumptions the function
$\cU(x,t)$ defined by eq.~\pref{shortdistcorrs2} has dimension
(mass)${}^{2d-4}$. For later numerical estimates there are two
cases of interest: ($i$) a dimension-4 (marginal) inflaton-bath
coupling, for which $d=3$ and $g$ is dimensionless; and ($ii$) a
dimension-5 interaction with $d=4$ and coupling $g = 1/M$, for
some large mass scale $M$.

Suppose now that the physics of sector $B$ is homogeneous in space
and is characterized by a single mass scale, $\Lambda(t)$, which
can be slowly evolving with time as the universe expands. For
instance, $\Lambda(t)$ might be given by the temperature $T(t)$ if
$B$ were described by a simple thermal state. Working in flat
space we see that on dimensional grounds we can take
\be \label{ScaleEstimate}
    \langle \cB(x,t) \rangle_B = c_1 \, \Lambda^d(t)
    \quad \hbox{and}
    \quad
    \cU(x,t) = c_2 \, \Lambda^{2d-4}(t) \,,
\ee
where $c_1$ and $c_2$ are calculable dimensionless real numbers
(which might also depend logarithmically on $\Lambda$, for large
$\Lambda$). We imagine $c_2 \ge 0$, which should be true for a
broad class of choices for the environment (sector $B$).

Using these expressions, we then find
\be
    \langle V(t) \rangle_B = g c_1 \, \Lambda^d(t)
    \, a^3(t) \int d^3x \; v(x,t) \,,
\ee
and
\bea \label{dmeqscalar}
    \frac{\partial \rho_A}{\partial t} &=& -i \Bigl[
    \langle V \rangle_B ,
    \rho_A \Bigr]  \\
    && \quad - g^2  \, c_2 \,\Lambda^{2d-4}(t)\,
    a^3(t) \int d^3x \; \Bigl[
    v(x), \Bigl[ v(x) , \rho_A \Bigr] \Bigr]  \,. \nonumber
\eea

Evaluating this last equation within field eigenstates, $|V(x)
\rangle_t$, and integrating over times short compared with
$H^{-1}$ allows the neglect of the $H_A$ terms in $\partial_t
\rho_t[V,\tilde{V}]$, leading to the integrated result
\be \label{offdiagscalar}
    \rho_t[V,\tilde{V}] \approx \Bigl| \Psi_{t_0}[V]
    \Bigr|^2 \, e^{-\Gamma -i \Sigma} \,,
\ee
with the decoherence rate governed by the quantity
\be \label{GaussianFormScalar}
    \Gamma = g^2 c_2 \int_{t_0}^t dt' \Lambda^{2d-4}(t')\, a^3(t')
    \int d^3z \Bigl[ V(z) - \tilde{V}(z) \Bigr]^2
    \,.
\ee

Eq.~\pref{offdiagscalar} has the form of a Gaussian functional of
$V(x) - \tilde{V}(x)$, whose interpretation can be estimated by
evaluating it for configurations which extend over co-moving
distances of order $L$ and whose amplitude is $V_0$. For such
configurations we have
\be
    P_t(V_0,\tilde{V}_0)
    \sim P_{t_0} \exp\left[-\, \frac{L^{3}
    (V_0 - \tilde{V}_0)^2}{\sigma^2}
    \right]  \,,
\ee
with a width, $\sigma$, which evolves with time according to
\bea
    \frac{1}{\sigma^2} &=& g^2 c_2
    \int^t_{t_0} dt' \; \Lambda^{2d-4}(t') \, a^{3}(t') \nonumber\\
    &=& g^2 c_2\int^a_{a_0} \frac{\hat{a}^2 d\hat{a}}{H(\hat{a})} \;
    \Lambda^{2d-4}(\hat{a})
    \,.
\eea

To proceed we assume that $\Lambda$ evolves with time as does
temperature. Then $\Lambda(a) = \Lambda_0(a_0/a)$ and so the width
of the Gaussian density matrix acquires the time evolution
\bea
    \frac{1}{\sigma^2} &=& g^2 c_2\, \Lambda_0^{2d-4} a_0^3
    \int^{a/a_0}_{1} dx \; \left[ \frac{x^{6-2d}}{H(x)} \right] \nonumber\\
    &=& \frac{1}{9-2d} \left( \frac{g^2 c_2 \, \Lambda_0^{2d-4} a_0^3}{H_0}
    \right) \left[ \left(
    \frac{a}{a_0} \right)^{9-2d} - 1 \right]\,
\eea
where the last relationship assumes $H(a) = H_0 (a_0/a)^2$, as is
appropriate for a radiation-dominated universe.

The time-evolution of $\sigma$ differs qualitatively depending on
whether $d \ge 5$ or $d \le 4$. If $d \ge 5$ (corresponding to a
dimension-6 or higher inflaton-matter interaction) then
$1/\sigma^2$ saturates to $g^2 c_2 \, \Lambda_0^{2d-4}
a_0^3/[(2d-9)H_0]$ as $a/a_0$ grows, indicating that for
inflaton-matter couplings this weak the Gaussian maintains an
essentially fixed width as the universe expands. By contrast, for
$d \le 4$ we find $\sigma$ shrinks to zero as $a/a_0$ grows
without bound, indicating that couplings this strong lead at late
times to a diagonal distribution in the $|V\rangle$ basis,
corresponding to a stochastic statistical ensemble of classical
field configurations, $V(x)$.

For the two cases of most interest (a dimension-5 inflaton-matter
interaction with $d=4$ and $g = 1/M$, or a dimension-4 interaction
with $d=3$ and $g$ dimensionless) we find
\bea
    \frac{L^3}{\sigma^{2}} &=& \left( \frac{\Lambda_0^{4} \, \ell_0^3}{M^2 H_0}
    \right) \left( \frac{a}{a_0} \right) \quad \hbox{(if $d=4$)}
    \nonumber\\
    &=& \left( \frac{g^2 \Lambda_0^{2} \, \ell^3(a)}{3H_0}
    \right) \quad \hbox{(if $d=3$)} \,,
\eea
for large $a/a_0$, where $\ell(a) = L \, a$ denotes the physical
scale corresponding to the co-moving length $L$. These expressions
show that $\sigma$ shrinks like $(a/a_0)^{-1/2}$ or
$(a/a_0)^{-3/2}$ in these two cases as the universe expands.

We may also use these to estimate whether sufficient time can pass
to decohere the modes of interest for CMB observations. To this
end we evaluate the Gaussian distribution at the epoch of
radiation-matter equality, for modes with amplitude $V_0 -
\tilde{V}_0 \sim H_{\rm eq} = H(a_{\rm eq})$ and physical extent
$\ell^{-1}(a_{\rm eq}) \sim H_{\rm eq}$, starting with an initial
pure state at the reheat epoch (for which $a_0 = a_{\rm rh}$ and
$H_0 = H_{\rm rh} \sim T_{\rm rh}^2/M_p$). Using $H_{\rm eq} \sim
T_{\rm eq}^2/M_p$, $a_{\rm eq}/a_0 = T_{\rm rh}/T_{\rm eq}$,
$\ell_0 = \ell(a_{\rm eq}) (a_0/a_{\rm eq}) = H_{\rm
eq}^{-1}(a_0/a_{\rm eq})$ and $H_0 = H_{\rm eq} (a_{\rm
eq}/a_0)^2$ we have
\bea
    \left( \frac{L^3 H_{\rm eq}^2}{\sigma^{2}}
    \right)_{a=a_{\rm eq}} &=& \left(
    \frac{\Lambda_0^{4}}{M^2 H_{\rm eq}^2}
    \right) \left( \frac{T_{\rm eq}}{T_{\rm rh}}
    \right)^4 \sim \frac{M_p^2}{M^2}
    \qquad \hbox{(if $d=4$)}
\eea
and
\bea
    \left( \frac{L^3 H_{\rm eq}^2}{\sigma^{2}}
    \right)_{a = a_{\rm eq}}  &=&
    \left( \frac{g^2 \Lambda_0^{2}}{3H_{\rm eq}^2}
    \right) \left( \frac{T_{\rm eq}}{T_{\rm rh}}
    \right)^2 \sim \frac{g^2 M_p^2}{T_{\rm eq}^2}
    \qquad \hbox{(if $d=3$)} \,,
\eea
where the last approximate equalities use $\Lambda_0 \sim T_{\rm
rh}$.

Adequate decoherence requires $L^3 (V - \tilde{V})^2 / \sigma^2
\gg 1$, and so we see that for $d=3$ this is true at
radiation-matter equality for any $V - \tilde{V} \sim H_{\rm eq}$
given any reasonable choice for the dimensionless coupling $g$. It
also occurs when $d=4$ provided only that the scale $M$ associated
with the inflaton-matter coupling is smaller than the Planck
scale. This shows that modes whose amplitudes differ in amplitude
by as little as $H_{\rm eq}$ behave as if they have decohered by
the time of radiation-matter equality.

Alternatively, we can ask for what epoch does the combination
$L^3(V - \tilde{V})^2/\sigma^2$ become $O(1)$ for a mode with $V -
\tilde{V} \sim H_{\rm eq}$ and $\ell(a_{\rm eq}) \sim H^{-1}_{\rm
eq}$. Using the above expressions we see this occurs when
\bea
    \frac{a}{a_0} &\sim&
    \frac{M^2 \, T_{\rm rh}}{M_p^2 \, T_{\rm eq}}
    \quad \hbox{(if $d=4$)} \nonumber\\
    \frac{a}{a_0} &\sim&
    \frac{T_{\rm rh}}{(g^2 M_p^2 \, T_{\rm eq})^{1/3}}
    \quad \hbox{(if $d=3$)} \,.
\eea
If we demand that such modes decohere not long after the reheating
epoch, such as when $a/a_0 \sim N$ for $N$ not too much bigger
than 1, then this requires
\bea
    T_{\rm rh} &\sim&
    \left( \frac{N M_p^2  \, T_{\rm eq}}{M^2}
    \right) \sim 10^6 N \, \hbox{GeV} \left( \frac{10^{10} \,
    \hbox{GeV}}{M} \right)^2
    \qquad \hbox{(if $d=4$)} \,,
\eea
and
\be
    T_{\rm rh} \sim 10^9 N g^2  \, \hbox{GeV}
    \quad \hbox{(if $d=3$)} \,.
\ee
For reheat temperatures of this order it is not necessary to trust
the solution, eq.~\pref{offdiagscalar}, for too long.

Finally, notice that if $M \sim M_p$ (when $d=4$) then we have
$L^3 (V - \tilde{V})^2/ \sigma = O(1)$ for $V - \tilde{V} \sim
H_{\rm eq}$ regardless of the reheat temperature, and so in this
case all modes which differ in amplitude by an amount of order
$H_{\rm eq}$ would be just beginning to decohere from one another
as they re-enter the horizon. It is tempting to speculate that
this might have measurable implications for the observed
temperature fluctuations in the CMB.

\subsection{Decoherence During Inflation?}

We next explore the extent to which short-wavelength modes can
decohere longer-wavelength modes during inflation itself, before a
mixed environmental state has appeared. Our goal in this section
is twofold. We first show why decoherence within this epoch cannot
be reliably computed using the master-equation techniques
presented within this paper --- typically because of the failure
of the short-time correlation assumptions,
eqs.~\pref{shorttimecorrs} and \pref{shortdistcorrs}. We then
argue why it should be unlikely that short-wavelength modes can
decohere long-wavelength ones once a reliable method of
calculation becomes possible.

\subsubsection{Obstructions to Calculation}

Since our interest for observations is in those modes having
observational implications for the CMB, we again ask that sector
$A$ consist of those long-wavelength modes of the Mukhanov field,
$v$, which satisfy $k/a_{\rm he} \sim H$ at the epoch of horizon
exit. All other modes of the field $v$ are not observed, and so
can be lumped together and included into sector $B$ for the
purposes of CMB observations. To this end it is useful to divide
the field $v$ into a short- and long-wavelength part, $v = v_l +
v_s$, where the long-wavelength modes, $v_l$, are those modes
whose decoherence we wish to follow (sector $A$) and the
short-wavelength modes, $v_s$, satisfy $k/a_{\rm he} \gg H$ and
provide part of the environment over which we trace. (We ignore
those modes of $v$ whose wavelengths are much longer than those
which are observed, since their influence cannot be computed
reliably using our methods of calculation.)

Since the environment satisfies $k/a_{\rm he} \gg H$, these modes
start off in the ground state, $|0 \rangle$, until a later epoch
arises for which $k/a \sim H$, after which they start to become
squeezed and so approach their Bunch Davies vacuum, $|B \rangle$.
Of course, the environment (sector $B$) also includes the modes of
all of the many other fields which are present in the high-energy
theory besides the inflaton, many of which have masses $M \gsim H$
and so for which all modes may be regarded as being `high-energy'
in comparison with the observable inflaton modes. For the present
purposes we assume all of these modes to be in their Bunch-Davies
vacuum state, which does not differ appreciably from the Minkowski
vacuum to the extent that $k/a \gg H$ and/or $M \gg H$.

Consider, then, the following interaction lagrangian density
\be
    {\cal L}_{\rm int} = - g \, \cB[v_h, \psi] \, v_l^p\,,
\ee
where $v_l$ denotes as above the observable inflaton modes, $p$ is
some positive-integer power and $\cB$ is some local functional of
the inflaton's short-wavelength modes, $v_h$, as well as of all
other heavy fields, $\psi$. For instance, for the
self-interactions of the inflaton we can use the estimate provided
in Appendix \ref{App:MIPerturbations} for the strength of its
lowest-dimension self-interaction,
\be
    {\cal L}_{\rm int} = - \xi \, v^3 = - \xi (v_h^3 +
    3 v_h^2 v_l + 3 v_h v_l^2 + v_l^3) \,,
\ee
with $\xi$ dominated (for a simple model of chaotic inflation) by
the gravitational self-interactions, with strength
\be
    \xi \approx \pm\,\frac{48\, a H^2}{(2 \epsilon)^{3/2} M_p} \,.
\ee
Here $H$ is the inflationary Hubble scale, and $\epsilon$ is the
standard inflationary slow-roll parameter, $\epsilon = \frac12
(V'/V)^2$. Since for a cubic interaction momentum conservation
allows couplings between any three modes for which the sum $(k_1 +
k_2 + k_3)^\mu = 0$, we see that cubic interactions can only
couple very-short to very-long wavelength modes if two of the
modes have short wavelengths and one has long. That is, only the
term $-3\,\xi \, v_h^2 v_l$ which can have nontrivial matrix
elements and so participate in the mixing of very-long and
very-short wavelength modes.

To compute the influence of sector $B$ for the decoherence of
$v_l$ we must first estimate the relevant correlation function,
$\langle \delta \cB(x,t) \delta \cB(x',t') \rangle$, and see
whether it satisfies either of the short-correlation assumptions,
eqs.~\pref{shorttimecorrs} and \pref{shortdistcorrs}. We must do
so with the environment described by a pure squeezed state, which
rapidly approaches the attractor corresponding to the Bunch-Davies
vacuum, $|B \rangle$. For instance,starting with cubic inflaton
self-interactions leads to the correlation function $\langle
B|\delta v_s^2(x,t) \, \delta v_s^2(x',t') |B \rangle$, while an
inflaton/heavy-field interaction of the form ${\cal L}_{\rm int} =
g \psi \, v^p$ for a field with $\langle B | \psi | B \rangle = 0$
would simply require the calculation of the Wightman function,
$\langle B | \psi(x,t) \psi(x',t') | B \rangle$.

The main observation to be made is that when such a correlator is
computed using the Bunch Davies vacuum, it generically does not
satisfy the condition eq.~\pref{shortdistcorrs}. This is most
easily seen for a heavy scalar field, $\psi$, of mass $m$, for
which the calculation has been made explicitly \cite{dSWF} (see
also \cite{dSProps}), with the result (in conformal time)
\be \label{BDWightman}
    \langle B | \psi(x,\eta) \psi(x',\eta') | B \rangle =
    \left( \frac14 - \nu^2 \right) \frac{H^2 \sec( \pi \nu)}{16
    \pi} \, F\left[ \frac32 + \nu, \frac32 - \nu, 2 ; 1 +
    \frac{s^2}{4 \eta \eta'} \right] \,,
\ee
where $s^2(x,\eta; x',\eta') = (\eta-\eta' - i \epsilon)^2 -
\delta_{jk} (x-x')^j (x-x')^k$, and $F[a,b,c;z]$ denotes the
standard hypergeometric function. The complex parameter, $\nu$, is
defined by $\nu^2 = \frac94 - m^2/H^2$, and becomes pure imaginary
in the limit $m > \frac32\, H$. This expression does not have the
$\delta$-correlated form of eq.~\pref{shortdistcorrs} even in the
limit $m \gg H$, because of the dependence on the separations in
space and time only through the combination $s(x,\eta; x',\eta')$.
The symmetries of de Sitter space ensure that this property is
quite general for the correlations of scalar fields within their
Bunch-Davies vacuum.

One might hope to press on despite of this by returning to first
principles and simply using an expression like
eq.~\pref{BDWightman} in the general result,
eq.~\pref{masterequation}, of Appendix \ref{App: ABFormalism}.
However since this equation is derived in perturbation theory, the
result obtained for $\rho_A(t)$ in this way by integration
generally cannot be trusted for times satisfying $Vt \gsim O(1)$.
For most applications this ensures that the result obtained cannot
be used for any times of cosmological interest, reflecting the
generic buildup of complicated correlations between sectors $A$
and $B$ as the sytem evolves. It is only because the short
correlation-time assumption --- $V\tau \ll 1$, or equivalently
eq.~\pref{shortdistcorrs} --- makes the time-evolution problem
into a Markov process that it is possible to trust the solutions
to eq.~\pref{masterequationbody} for $Vt \gg 1$.

\subsubsection{Why Decoherence During Inflation is Unlikely}

Although we see that it is difficult to compute decoherence
effects during inflation within a controllable approximation,
there are nevertheless strong reasons for doubting that
interactions with short-wavelength field modes during inflation
can decohere long-wavelength field modes. The reason hinges on the
assumption that these short-wavelength modes are themselves in
pure states, which in practice are well-approximated by the
Bunch-Davies vacuum. This assumption is certainly a natural one to
make for inflaton modes, given that the Bunch-Davies state is an
attractor of the squeezing equations \cite{Albrecht:1992kf}. It is
equally natural for the modes of other heavy fields (although it
can be violated for some choices of scalar-field initial
conditions \cite{TPI}).

The point is that for short-wavelength modes the Bunch-Davies
state is very close to the ordinary Minkowski vacuum, and so any
decoherence which is generated by a short-wavelength mode during
inflation is also likely to be present in flat space, in the
absence of inflation. If it were true that having short-wavelength
modes in their vacuum sufficed to decohere long-wavelength modes,
it would be necessary to understand why we do not see the vacuum
decohere around us all the time, and why quantum coherence is
possible at all for the many low-energy states which have been
studied throughout physics. Any convincing demonstration of
decoherence during inflation by tracing out pure-state,
short-wavelength modes must also explain why the same effect does
not decohere long-wavelength modes within Minkowski space.

\section{Discussion}

In this paper we present a formalism within which it is possible
to follow explicitly the development of decoherence within a
system's long-wavelength modes due to their interactions with
various kinds of short-wavelength physics. The formalism relies
for its validity on there being a large hierarchy between the
correlation-time for fluctuations in the short-distance sector and
the time-frame for evolution of the long-wavelength modes.

By applying this formalism to several possible choices for the
short-distance environment we show that there is ample time during
inflation for decoherence to occur between reheating and horizon
re-entry for the modes of interest for CMB observations. We
emphasize that our calculations do not yet establish what the most
important source of decoherence might be for inflationary
perturbations. In particular, we cannot compute with these tools
how decoherence proceeds due to environments whose correlation
times are not much shorter than the Hubble scale, or for
short-wavelength modes prepared within their Bunch-Davies vacuum
during inflation itself, although, we argue that decoherence during
inflation by short-wavelength modes in their vacuum is unlikely,
given the absence of such decoherence in non-cosmological
situations.

Rather, our intent is merely to show that plausible sources could
easily have sufficed to decohere the observed modes.
Interestingly, if the decoherence is due to gravitational-strength
interactions with the thermal (or other mixed) state produced
during reheating, it could well be that the process is only just
occurring for modes whose amplitudes differ by $O(H)$ as they
re-enter the horizon near radiation-matter equality. If this
should prove to be the most important source of decoherence we
believe it would be worth exploring whether this might have
observable consequences.

We argue that the decohered system tends to a final density matrix
which is diagonal in a basis which also diagonalizes the relevant
interaction Hamiltonian. On grounds of locality this is typically
a basis for which the long-distance field operators themselves are
diagonal. Since such a density matrix corresponds physically to
the establishment of a stochastic distribution of classical field
configurations, it is precisely the kind of late-time state which
is implicitly assumed in standard calculations of the impact of
inflationary fluctuations on the CMB.

\medskip\begin{center} {\it Note Added:} \end{center}

While preparing this manuscript we discovered a paper,
ref.~\cite{pm}, posted to the arXiv by one of our collaborators on
this work. It describes an analysis which overlaps the one
presented here on the calculation of the effects of decoherence by
squeezed states during inflation. We believe that it differs in
its conclusions due to errors in the calculations presented there,
including (but not restricted to) the application of the
master-equation formalism outside of its domain of validity.

\section*{Acknowledgements}

We thank Robert Brandenberger, Shanta de Alwys and Mark Sutton for
fruitful discussions which helped to shape our ideas on this
subject, and Patrick Martineau, an ex-student to whom the
application of this formalism was assigned as a thesis problem.
C.B. and R.H. are grateful to the Aspen Center for Physics for
providing such a pleasant environment within which some of this
research was performed. R.~H. was supported in part by DOE grant
DE-FG03-91-ER40682, while the research of C.B. and D.H. is
partially supported by grants from N.S.E.R.C. (Canada), the Killam
Foundation, McGill and McMaster Universities and the Perimeter
Institute.

\appendix

\section{Interactions With a Fluctuating Environment}
\label{App: ABFormalism}

In this Appendix we summarize the formalism we use to determine
the evolution of the density matrix for slow degrees of freedom
interacting with an environment which contains fast fluctuations.
We keep our exposition general, specializing to the case of
interest at the end of the section.

\subsection{General Formulation of the Problem}

Our goal is to provide a general master equation which dictates
the evolution of the density matrix of a system, $A$, of slow
degrees of freedom which interacts with an environment, $B$, of
faster modes. We do not follow any of the $B$ observables, and
imagine that any correlations which the $B$ variables induce into
the $A$ variables have correlation times which are very short
compared with the time scale of the $A$-physics which is of
interest. To this end we use a formalism which was developed for
applications to physics in condensed-matter and atomic systems
\cite{API,NOptics,formalism}, but which has also seen service in
describing neutrino evolution within complicated astrophysical
environments \cite{BM}.

Specifically, we imagine the system's Hilbert space to be the
direct product of the $A$ and $B$ Hilbert spaces, ${\cal H} =
{\cal H}_A \otimes {\cal H}_B$, and the total Hamiltonian to be
described by the following sum
\be
    H = H_0 + V = H_A \otimes I + I \otimes H_B + V \,,
\ee
where $H_A$ and $H_B$ are the Hamiltonians of the subspaces $A$
and $B$ in the absence of any interactions, and $V$ is the
interaction term. In future we often suppress the presence of the
unit operators, $I$. We also imagine the system starts off being
described by a density matrix, $\rho$, for which there are no
initial correlations between $A$ and $B$: $\rho(t = t_{0}) =
\varrho_A \otimes \varrho_B$.

In these circumstances a mean-field expansion allows practical
progress, assuming that the correlation times, $\tau$, are small
enough that $V \tau \ll 1$ for the matrix elements of interest.
The mean-field expansion must be defined in terms of the
time-evolution operator, since only in this case do the mean-field
and fluctuation parts not interfere with one another. That is, if
we consider the time-evolution operator in the interaction
representation, then
\be \rho(t) = U(t) \, \rho(t = t_{0}) \, U^*(t) \ee
with
\be
    \frac{\partial U(t)}{\partial t} = - i V(t) \, U(t) \,,
\ee
where $V(t) = e^{i H_0 t} \, V \, e^{-iH_0t}$.

Let us define the mean field $\overline{U}$ by
\be
    \overline{U} = \langle U \rangle_B = \Tr_B[ \rho U ] \,,
\ee
where the trace is only over the states in sector $B$. Thus,
$\overline{U}$ is an operator which acts in ${\cal H}_A$. Defining
the fluctuation by $\Delta U = U - \overline{U}$, the exact time
dependence of any $A$-sector observable, ${\cal O}_A$, can be
written as
\bea
    \langle {\cal O}_A \rangle (t) &=& \Tr \Bigl[ \rho(t) \, {\cal O}_A
    \Bigr]
    \nonumber \\
    &=& \Tr\Bigl[ U(t) \, \rho(t=t_{0}) \, U^*(t) \,
        {\cal O}_A \Bigr] \label{last} \\
    &=& \Tr_A \Bigl[ \overline{U}(t) \, \rho_A(t = t_{0}) \,
        \overline{U}^*(t)
    \, {\cal O}_A \Bigr] \nonumber\\
    && \qquad + \Tr \Bigl[ \Delta U(t) \, \rho(t = t_{0})
        \, \Delta U^*(t) \, {\cal O}_A \Bigr] \,. \nonumber
\eea
In the first term of the last line of this equation, $\rho_A$ is
the reduced density matrix
\be \rho_A(t) = \Tr_B[\rho(t)] \,, \ee
which suffices to follow the time-dependence of any measurement
only performed in sector $A$.

The last line of (\ref{last}) also shows that defining the mean in
terms of $U$ (rather than, say, $H$) guarantees that the time
evolution of all observables decomposes into the sum of a mean
evolution plus a fluctuation about this mean, with no cross terms
involving both $\overline{U}$ and $\Delta U$. Indeed, this is the
point of using $U$ (rather than $V$, say) when making the split
between mean and fluctuating variables.

We now compute the differential evolution of $\rho_A(t)$ using
perturbation theory, with $V$ being the perturbation. Expanding
the exact evolution equation gives
\bea \label{masterequation}
    &&\frac{\partial \rho_A}{\partial t}(t) = -i \Bigl[
    \overline{V}(t) \, \rho_A(t) - \rho_A(t) \,
    \overline{V}^*(t) \Bigr] \\
    && \qquad + \int_{t_{0}}^t d\tau \, \Tr_B \Bigl[  \delta V(t)
    \, \rho(t) \, \delta V(\tau) + \delta V(\tau) \, \rho(t) \,
    \delta V(t) \Bigr] \nonumber \\
    && \qquad + O(V^3) \,. \nonumber
\eea
In these expressions $\delta V = V - \overline{V}$ and
$\overline{V}$ is defined in terms of $\overline{U}$ by
\be
    \overline{V}(t) = \left( i \frac{\partial \overline{U}}{\partial t}
    \right) \, \overline{U}^{-1} \,,
\ee
so
\be
 \overline{V}(t) =
\langle V \rangle_B -i \int_{t_{0}}^t d\tau \, \langle
    \delta V(t) \delta V(\tau) \rangle_B + O(V^3) \,.
\ee

Notice that $\overline{U}$ need not be unitary, because
probability can scatter out of sector $A$ into sector $B$, or vice
versa. Consequently, $\overline{V}$ need not be hermitian.
However, an important special case for which $\overline{V}$ {\it
is} unitary arises when the $B$ sector involves only states which
are too heavy to be produced by the energy available in the $A$
sector, and the $B$ sector is prepared in its vacuum state:
$\rho_B = |0\rangle_B {}_B\langle 0|$. The evolution of states in
sector $A$ is guaranteed to be unitary in this case because there
is no energy stored for release in sector $B$, and there is
insufficient energy available in sector $A$ to excite states in
sector $B$, and to thereby complicate the density matrix in this
sector.

In the event that the system $B$ is large enough that its state is
largely unchanged by the interactions with $A$, we can evaluate
the right hand side of eq.~(\ref{masterequation}) in terms of
$\rho_A(t)$ by writing $\rho(t) \approx \rho_A(t) \otimes
\varrho_B$, and thereby set up a differential equation which may
be integrated to obtain the long-time behavior of $\rho_A(t)$. The
point of this formulation is that this evolution can be trusted
even over very long time scales, provided that the correlation
time of system $B$ is sufficiently short.

Similarly, if we write the interaction Hamiltonian in the form
$V(t) = A_j(t) \otimes B_j(t)$ (with an implied sum over `$j$'),
and assume that the correlations in the fluctuations of $B_j$ are
short-lived,
\be
    \langle \delta B_j(t) \delta B_k(t') \rangle_B \equiv
    \Tr_B \Bigl[ \varrho_B \delta B_j(t) \delta B_k(t') \Bigr]
    = {\cal W}_{jk}(t) \delta(t-t') \,,
\ee
then we find
\bea
    \overline{V}(t) &=& A_j(t) \langle B_j(t) \rangle_B -
    \frac{i}{2} \, {\cal W}_{jk}(t) A_j(t) A_k(t) \nn\\
    \hbox{and} \quad
    \overline{V}^*(t)  &=& A_j(t) \langle B_j(t) \rangle_B +
    \frac{i}{2} \, {\cal W}_{jk}(t) A_j(t) A_k(t) \,.
\eea
These expressions use $\int_{t_0}^t dt' \; f(t') \delta(t-t') =
\frac12 \, f(t)$ and ${\cal W}^*_{jk}(t) = {\cal W}_{kj}(t)$. The
master evolution equation for $\rho_A(t)$ then is (neglecting
$O(V^3)$ terms)
\be \label{masterequation2}
    \frac{\partial \rho_A}{\partial t} = i \Bigl[\rho_A
    , A_j \Bigr] \langle B_j \rangle_B
    -\, \frac12 \, {\cal W}_{jk}  \Bigl\{ \Bigl(
    \rho_A A_j A_k + A_j A_k \rho_A - 2 A_k \rho_A A_j
    \Bigr) \Bigr\} \,.
\ee

\subsection{Microscopic Fluctuations}

We now specialize to the case where the degrees of freedom in
sector $B$ also have very short wavelength, in which case their
influence can be represented in terms of local interactions in the
spirit of effective field theories. In this case the constraints
of locality allow considerable simplification. Suppose, then, that
sector $A$ and $B$ interact through a local interaction of the
form:
\be \label{VdefApp}
    V = \int d^3x \; \Bigl[ \cA_j(x,t) \otimes \cB_j(x,t)
    \Bigr] \,,
\ee
where ${\cal A}_j$ denotes an ultra-local function of the fields
describing the $A$ degrees of freedom, and ${\cal B}_j$ plays a
similar role for sector $B$.

In terms of this interaction, the interaction potential,
$\overline{V}$, is expressible (using the formulae given above) in
terms of correlations of the variables ${\cal B}_j$. Using the
assumption that these correlations in the $B$ sector are very
short compared with the other scales of interest, we may write
\be \label{LocalCorrelation}
    \langle \delta {\cal B}_j(x,t) \, \delta {\cal B}_k(x',t')
    \rangle_B = \cU_{jk}(x,t) \, \delta^3(x-x') \, \delta(t-t') \,,
\ee
where the $\cU_{jk}(x,t) = \cU^*_{kj}(x,t)$ are calculable local
functions of position. Given this assumption, we have
\be
    \int_{t_{0}}^t dt' \,
    \langle \delta V(t) \delta V(t') \rangle_B =
    \frac12 \int d^3x \,  {\cal A}_j {\cal A}_k
    \, \cU_{jk}(x,t)  \,.
\ee
and so we see that $\overline{V}(t) = \int d^3x \,
\overline{\cV}(x,t)$, with
\be \label{VbarDefn}
    \overline{\cV} = {\cal A}_j \langle {\cal B}_j(x,t)
    \rangle_B - \frac{i}{2} \cA_j \cA_k \, \cU_{jk}(x,t) + O(V^3)
    \,.
\ee
If probability and energy transfer can only flow into sector $B$
from sector $A$, then the matrix function $\cU_{jk}$ is
non-negative definite.

Using these expressions for the correlations of the energy density
in the environment, we can write the master equation for
$\rho_A(t)$ in the following way, accurate to second order in the
coupling:
\be \label{dmeq}
    \frac{\partial \rho_A}{\partial t} =
    i  \int
    d^3x \; \Bigl[ \rho_A , \cA_j
    \Bigr] \, \langle \cB_j
    \rangle_B
    -  \frac12 \int d^3x \; \cU_{jk} \Bigl[
    \cA_j \cA_k \rho_A + \rho_A \cA_j \cA_k
    - 2 \cA_k \rho_A \cA_j \Bigr]  \,.
\ee
This is the result quoted as eq.~\pref{dmeqbody} in the main text.

\subsubsection{Example: Photons in a Thermal Fluid}

An explicit example of practical interest to which the key
assumption, eq.~\pref{LocalCorrelation}, applies is that of light
propagating through an electrically neutral medium, for which
sector $A$ consists of the photons and sector $B$ is the medium.
Taking the photon-medium interaction to be given by
\be
    {\cal L}_{\rm int} = e A_\mu  J^\mu \,,
\ee
where $J^\mu$ is the electric-current operator, shows that we may
take $\cA = e A_\mu$ and $\cB = J^\mu$. Provided the medium is
electrically neutral and carries no net currents we then find
$\langle J^\mu \rangle_B = 0$, and eq.~\pref{LocalCorrelation}
becomes
\be
    \langle J^\mu(x,t) J^\nu(x',t') \rangle_B \approx \Pi^{\mu\nu}(x,t) \,
    \delta^3(x - x') \delta(t-t') \,,
\ee
where we follow conventional practice and use the notation
$\Pi^{\mu\nu}$ (instead of ${\cal U}_{jk}$) for the current
correlation function.

With these assumptions the effective photon lagrangian,
eq.~\pref{VbarDefn} becomes
\be
    \overline\cV = - \frac{ie^2}{2}  A_\mu A_\nu \Pi^{\mu\nu}(x,t)
    + O(V^3) \,,
\ee
which shows that photon propagation through a medium is governed
by the medium's polarization tensor, $\Pi^{\mu\nu}$, with photon
scattering being governed by fluctuations of the electric current
within the medium. In the special case where the medium's current
fluctuations are due to thermal fluctuations in a
rotationally-invariant fluid which is in local thermal
equilibrium, then they have the form
\be
    \Pi^{\mu\nu} = \left( \begin{array}{cc}
    \pi_1 & 0 \\ 0 & \pi_2 \, \delta^{ij} \end{array} \right) \,,
\ee
with local quantities, $\pi_1$ and $\pi_2$, which are calculable
in terms of the local thermodynamic variables. If the temperature
is much higher than the other relevant scales then we expect on
dimensional grounds that $\pi_a = k_a \, T^2$, where $k_1$ and
$k_2$ are calculable dimensionless constants, leading to a photon
scattering rate which is of order $\Gamma \sim e^2 T^2$.
Ref.~\cite{BM} shows that a similar analysis for neutrino
propagation reproduces the standard results for neutrino
scattering by a thermal bath.

\subsection{Pure-to-Mixed Transitions}

We next identify what features of the master equation play a role
in making transitions from pure states to mixed states.

A pure state is defined by the condition that there exists a state
vector, $|\psi\rangle$, for which the density matrix can be
written $\rho_A = |\psi \rangle \langle \psi |$. This is a special
case of the diagonal representation for $\rho_A$,
\be
    \rho_A = \sum_n p_n |n \rangle \langle n | \,,
\ee
for which $p_n = 0$ for all $n \ne \psi$, and $p_\psi = 1$ for $n
= \psi$. This is equivalent to the basis-independent condition
$\rho_A^2 = \rho_A$. Since $\rho_A$ is non-negative definite and
initially normalized, $\Tr[\rho_A] = 1$, the coefficients $p_n$
satisfy $0 \le p_n \le 1$ and so may be regarded as probabilities.
In particular this implies $p_n^2 \le p_n$, with $p_n^2 = p_n$ if
and only if $p_n = 0,1$. It follows that $\rho_A$ describes a pure
state if and only if $\Tr[\rho^2_A] = \Tr[\rho_A] = 1$.

Motivated by these observations, we next use the master evolution
equation, \pref{masterequation2}, to evaluate how $\Tr[\rho_A]$
and $\Tr[\rho_A^2]$ vary with time. As is simple to see,
\bea
    \frac{\partial}{\partial t}\, \Tr[\rho_A] &=& i
    \langle B_j \rangle_B \Tr\Bigl[\rho_A
    , A_j \Bigr]
    -\, \frac12 \, {\cal W}_{jk} \Tr \Bigl\{ \Bigl(
    \rho_A A_j A_k + A_j A_k \rho_A - 2 A_k \rho_A A_j
    \Bigr) \Bigr\} \nn\\
    &=& 0\,,
\eea
and the rate for the pure-to-mixed transition is
\bea
    \frac{\partial}{\partial t} \, \Tr[\rho^2_A] &=&
    \Tr\left[ \rho_A \frac{\partial \rho_A}{\partial t} +
    \frac{\partial \rho_A}{\partial t} \rho_A \right] \nn\\
    &=& \Bigl( {\cal W}_{jk} + {\cal W}_{kj} \Bigr)
    \,  \Tr \Bigl[ \rho_A A_j \rho_A A_k \Bigr]
    - 2 {\cal W}_{jk} \Tr \Bigl[ \rho_A^2 A_j A_k \Bigr]
    \,.
\eea
Evaluating this for an initially pure state, $\rho_A(t_0) = |
\alpha \rangle \langle \alpha |$, then leads to
\be
    \left. \frac{\partial}{\partial t} \, \Tr[\rho^2_A]
    \right|_{t_0} =
    -2 {\cal W}_{jk}(t_0) \langle \alpha |
    \delta A_j \delta A_k | \alpha \rangle
    \,,
\ee
where $\delta A_j = A_j - \langle \alpha | A_j | \alpha \rangle$.

\section{Metric-Inflaton Perturbations}
\label{App:MIPerturbations}

Before turning to explicit choices for the environment, we first
pause to describe the observable sector (sector $A$) of interest
for inflationary calculations. This allows us to identify the
modes of interest for inflationary perturbations, and to estimate
the size of their self-couplings.

\subsection{Linearized Modes}

For inflationary applications the observable system consists of
those perturbations to the energy density which have wavelengths
corresponding to those which are responsible for the observed
temperature fluctuations in the CMB spectrum. Following standard
practice we take these to be described by the appropriate
combination of the scalar perturbations of the metric and the
modes of the inflaton field, $\phi$, whose slow roll is
responsible for the occurrence of inflation. Our description here
draws extensively from the treatment in
refs.~\cite{Mukhanov,Brandenberger}.

For single-field inflation the relevant dynamics is described by
the action (using rationalized Planck units $M_p^{-2} = 8 \pi G =
1$)
\be \label{action}
    S = -\int d^{4}x \, \sqrt{-g} \left[ \frac{1}{2} \,R
    + \frac{1}{2} \partial_\mu\phi \, \partial^\mu\phi + V(\phi)
    \right] \,.
\ee
Inflation can occur if the universe should enter a phase during
which the total energy density is dominated by a slowly-rolling
homogeneous expectation value, $\phi(t)$, whose kinetic energy is
much smaller than its potential energy: $\frac12 \, \dot\phi^2 \ll
V(\phi)$. This can occur -- given specific initial conditions --
if the inflaton potential, $V(\phi)$, is chosen to be sufficiently
flat. During inflation we take the space-time background to be
\be
    \exd s^2 =  - \exd t^2 + a^2(t) \, \gamma_{ij} \,
    \exd x^i \, \exd x^j \,,
\ee
where the Hubble scale, $H = \dot{a}/a$, is approximately constant
and $\gamma_{ij} = \delta_{ij}$ denotes the spatial metric, which
for simplicity we take to be flat.

The primordial fluctuations of relevance to the CMB within this
picture are the quantum fluctuations of the inflaton field and of
the scalar fluctuations of the metric. The wavelengths of these
fluctuations are stretched exponentially during inflation,
eventually to the size of the entire present-day observable
universe.

Once formed, the evolution of the fluctuation is traced by
linearizing the field equations about the inflationary background,
using $\phi = \varphi + \delta \phi$ and $g_{\mu\nu} =
\hat{g}_{\mu\nu} + \delta g_{\mu\nu}$, where $\varphi$ and
$\hat{g}_{\mu\nu}$ represent the background inflationary solution
to the equations of motion. Under these assumptions there is a
gauge for which the scalar fluctuations of the metric can be
written in terms of a scalar field, $\psi$, by
\be \label{eq:ptbedmetric}
    (\hat{g} +\delta g)_{\mu\nu} \, \exd x^\mu \exd x^\nu = a^2(\eta)
    \Bigl( -[1 + 2 \psi(x,\eta)] \, \exd \eta^{2}  +
    [1 - 2 \psi(x,\eta)] \, \gamma_{ij} \exd x^{i} \exd x^j
    \Bigr) \,.
\ee
We here switch to conformal time, defined by $\exd t = a(\eta) \,
\exd \eta$, and ignore vector and tensor metric fluctuations since
these are not yet relevant for CMB observations.

Only one combination of the two fields $\delta \phi$ and $\psi$
propagates independently, and within the linearized approximation
this combination is most efficiently identified in terms of the
variable \cite{Mukhanov2,Lukash}
\be \label{mukhvar}
    v =  a(\eta) \, \left[ \delta\phi +
    \frac{\varphi^{\prime}}{\cH}\psi \right] \,,
\ee
where primes denote differentiation with respect to $\eta$ and
$\cH = a'/a = H a$ is the conformal-time Hubble scale. Expanding
the action to quadratic order in $v$ gives:
\be \label{S_2}
    S_2\,=\,\frac{1}{2} \int{d^{4}x \, \left[ v^{\prime
    2}- \gamma^{ij} \partial_i v \, \partial_j v+
    \left( \frac{z^{\prime\prime}}{z}
    \right) v^{2} \right]},
\ee
where $z = a\varphi^{\prime}/\cH$. The utility of the variable $v$
follows because this action has canonical kinetic terms, and
simply describes the evolution of a free scalar field with a
time-dependent squared mass, $m^2 = - z^{\prime\prime}/z$,
propagating in flat spacetime. This makes its evolution and
quantization comparatively straightforward to understand.

Writing the fluctuation this way shows that the evolution of the
fluctuation amplitude depends crucially on the relative size of
the mode wave-vector, $k$, and the `mass' term, $-z''/z$. In
particular, this evolution changes qualitatively depending on
whether $k/a$ is larger or small than the Hubble scale $H$. What
is important is the relative size of these two changes during the
course of inflation and afterwards. During inflation the
wavelength, $\lambda_k(t) = a(t)/k$, stretches while the Hubble
length, $H^{-1}$ remains almost constant, and so the modes
relevant to the later universe at some point leave the horizon by
transiting to $\lambda > H^{-1}$ at the epoch of horizon exit,
roughly 60 $e$-foldings before the end of inflation. After
inflation the Hubble length grows faster than does $\lambda$, and
so modes begin to re-enter the Hubble radius once inflation ends
(see Fig.~\pref{Fig:Cartoon}). After re-entry these fluctuations
are imagined to provide the primordial seeds which are responsible
for the observed temperature fluctuations in the CMB. Between
horizon exit and re-entry, while $k/a \ll H$, the mode amplitude
satisfies $v_k'' \approx (z''/z)v_k$ and so $v_k(\eta) \propto
z(\eta) \propto a(\eta)$, corresponding to $\delta \phi$ being
approximately frozen. The proportionality $z \propto a$ holds if
the equation of state for the background geometry does not change
in time, as we henceforth assume.

The time-evolution of the quantum state of the field, $v$, may be
understood using the second-quantized Hamiltonian for the action
$S_2$, which is
\be \label{Hamilt}
    H_A = {H}_{\rm free} -i \left( \frac{z^{\prime\prime}}{z} \right)
    \int{d^{3}{k}}
    \Bigl( a_{k}\,a_{-k} - \hbox{\rm h.c.} \Bigr) \,,
\ee
where ${H}_{\rm free}$ is the usual harmonic-oscillator
Hamiltonian for each mode. Because the second term in this
expression is time-dependent, it causes the energy eigenstates to
evolve away from the adiabatic vacua, leading to the `squeezing'
for super-Hubble modes which satisfy $k/a \ll H$. Typically this
evolution leads to an attractor solution, corresponding to the
Bunch-Davies vacuum state. For sub-Hubble modes ($k/a \gg H$), by
contrast, the second term in (\ref{Hamilt}) is dominated by the
first term, and so this state does not differ much from the
`Minkowski' ground state defined by $a_k |0\rangle = 0$.

\subsection{Coupling Strengths}

For later use we next estimate the typical strength of the
self-interactions amongst these linearized fluctuations. It
suffices for our purposes to focus on cubic self-interactions of
the general form
\be
    {\cal L}_{\rm int} = -\xi \, v^3 \,,
\ee
for which we wish to estimate $\xi$. Since $v$ is a linear
combination of $\delta \phi$ and $\psi$, such couplings can
potentially arise either as inflaton or as gravitational
self-interactions. In this section we argue that the slow-roll
requirement on the inflaton potential ensures that gravitational
interactions provide the dominant cubic contribution.

\subsubsection*{Gravitational Self-Interactions}

An estimate the size of cubic metric self-couplings is obtained by
expanding the Einstein-Hilbert action to cubic order in $\psi$.
Evaluating the Ricci scalar for the metric of
eq.~\pref{eq:ptbedmetric} and dropping derivatives of $\psi$ leads
to
\be
    R = - \, \frac{ 6 \, a''}{a^3(1 + 2 \, \psi) \,
    }  + \hbox{($\partial \psi$ terms)}\,,
\ee
and so the cubic term in the Einstein-Hilbert lagrangian density,
${\cal L}_{EH} = -\frac12 \, \sqrt{-g} \, R$, is
\be \label{LgCubic1}
    {\cal L}_{3} = - 24 \, a a''  \psi^3
    = - \, \frac{24 a''}{a^2} \left( \frac{{\cal H}}{\varphi'} \right)^3
    v^3\,,
\ee
where the last equality expresses the result in terms of the
canonically-normalized variable obtained by dropping $\delta \phi$
from the definition of $v$: $\psi \sim [{\cal H}/(a\, \varphi')] v
= [H /(a\,\dot\varphi)] v$.

Specialized to the near-de Sitter geometry appropriate during
inflation we have $a''/a^2 = 2aH^2$ for constant $H$, related to
the scalar potential by $3H^2 \approx V$. Furthermore, ${\cal
H}/\varphi' = H/\dot\varphi \approx -3H^2/V' \approx \pm
(2\epsilon)^{-1/2}$, where $\epsilon = \frac12(V'/V)^2$ is the
usual slow-roll parameter and we use the slow-roll condition
$3H\dot\varphi \approx -V'$. (When applied to $V$, primes here
denote derivatives with respect to its argument, $\varphi$.)
Eq.~\pref{LgCubic1} then reduces to
\be
    {\cal L}_{3} = -48 \, a^4 H^2 \, \psi^3 = \mp \left[ \frac{48 a H^2}{
    (2\epsilon)^{3/2}} \right] v^3 \,,
\ee
where the upper (lower) sign applies if $\varphi' >0$ ($<0$).
Restoring $M_p = (8\pi G)^{-1/2}$, this represents a contribution
to the cubic coupling which during inflation is of order
\be
    \xi_g \approx \pm\,\frac{48\, a H^2}{(2 \epsilon)^{3/2} M_p} \,.
\ee

\subsubsection*{Inflaton Self-Couplings}

We next compare the estimate of the cubic gravitational
self-coupling just made with that obtained from the inflaton
scalar potential. Expanding $V$ to cubic order in $\delta \phi$
leads to the following cubic interaction ${\cal L}_3 = - \sqrt{-g}
\, V'''(\varphi) (\delta \phi)^3$. Since this depends on the third
derivative, its size depends somewhat on the kind of inflationary
potential which is used. For instance, for a simple model of
chaotic inflation, defined by a potential of the form $V = \lambda
\phi^{4}$, with $\lambda$ a dimensionless coupling constant, we
have
\be
    {\cal L}_3  = -4 \lambda  a^4 \varphi (\delta \phi)^3
    = -4 \lambda a \, \varphi \, v^{3} \,,
\ee
leading to the effective trilinear coupling constant $\xi_\phi = 4
\lambda a \,\varphi$.

How does this compare with the previously-computed gravitational
estimate? Taking the ratio $\xi_\phi/\xi_g$ gives
\be
    \frac{\xi_{\phi}}{\xi_{g}} = \frac{4 \,\lambda a\, \varphi \,
    }{\pm 48 a H^2 /[(2\epsilon)^{3/2} M_p]}
    = \pm \left[ \frac{\lambda M_p \, \varphi}{12H^2} \right]
    (2\epsilon)^{3/2}
    = \pm\, \frac{1}{4} \,
    \left[ \frac{\sqrt{2\epsilon} \, M_{p}}{\varphi} \right]^{3}
    = -\, \frac{\epsilon^3}{32}  \,,\\
\ee
where we use sign$\,\varphi' = -\,$sign$\,\varphi$, and simplify
using the expressions $3H^2 M_p^2 \approx V = \lambda \varphi^4$
and $2\epsilon = (V'/V)^2 = 16(M_p/\varphi)^2$ which are obtained
using the assumed inflaton potential. Since the existence of a
slow roll requires $\epsilon \ll 1$ we see that it is the graviton
self-coupling which dominates. In what follows we therefore use a
cubic self-interaction with strength given by the gravitational
self-coupling when computing decoherence.

\subsection{Squeezed States} \label{App: Squeezed States}

This Appendix briefly summarizes some of the useful properties of
the Bunch-Davies attractor, as described in the squeezed-state
language. Our description here follows that of
ref.~\cite{Albrecht:1992kf}.

Since the evolution of the field $v$ whose evolution is governed
by the Hamiltonian (\ref{Hamilt}), the evolution of its ground
state can be written $U |0\rangle$, where
\be
    U = S(r_{k} ,\Phi_k) \, R(\Theta_k) \,.
\ee
Here the operators $S$ and $R$ are defined by
\bea \label{squeezing}
    S(r_k,\Phi_k) &=& \exp\left[ \frac{r_k}{2} \left(
    e^{-2i\Phi_k} a_{-k} a_{k} - \hbox{h.c.} \right) \right],
    \nn\\
    R(\Theta_k) &=& \exp\left[ -i\Theta_k
    \Bigl( a^*_k a_k + a^*_{-k} a_{-k} \Bigr) \right] \,,
\eea
and $S(r_{k},\Phi_k)$ defines the state's squeezing, while
$R(\Theta_k)$ represents a rotation.

The parameters $r_k$, $\Phi_k$ and $\Theta_k$ are computable
functions of conformal time, $\eta$, given an explicit
inflationary scenario. An attractor solution for these parameters
is given by \cite{Albrecht:1992kf}:
\bea \label{deSittersqueeze}
    r_k &=& \sinh^{-1} \left( \frac{1}{2k\eta} \right) \nn\\
    \Phi_k &=& - \frac{\pi}{4} - \tan^{-1}\left( \frac{1}{2k\eta}
    \right) \\
    \Theta_k &=& k\eta + \tan^{-1}\left( \frac{1}{2k\eta} \right)
    \,,
\eea
corresponding to the Bunch-Davies vacuum of de Sitter space.

\end{document}